\documentclass[11pt,preprint]{aastex}
\usepackage{apjfonts}

\def\gax{\mathrel{\raise.3ex\hbox{$>$}\mkern-14mu\lower0.6ex\hbox{$\sim$}}}
\def\lax{\mathrel{\raise.3ex\hbox{$<$}\mkern-14mu\lower0.6ex\hbox{$\sim$}}}
\def\gtorder{\mathrel{\raise.3ex\hbox{$>$}\mkern-14mu
             \lower0.6ex\hbox{$\sim$}}}
\def\ltorder{\mathrel{\raise.3ex\hbox{$<$}\mkern-14mu
             \lower0.6ex\hbox{$\sim$}}}

\begin{document}

\title{Discovery of Variability of the Progenitor of SN~2011dh in M51 \\
   Using the Large Binocular Telescope}

\author{
   D.~M. Szczygie\l{}$^{1}$,
   J.~R. Gerke$^{1}$,
   C.~S. Kochanek$^{1,2}$, 
   K.~Z. Stanek$^{1,2}$
  }

\altaffiltext{1}{Department of Astronomy, The Ohio State University, 140 West 18th Avenue, Columbus OH 43210;
szczygiel, gerke, ckochanek, kstanek @astronomy.ohio-state.edu;
}
\altaffiltext{2}{Center for Cosmology and AstroParticle Physics, The Ohio State University, 191 W. Woodruff Avenue, Columbus OH 43210}

\begin{abstract}
We show that the candidate progenitor of the core-collapse SN~2011dh in M51
($8$~Mpc away) was fading by $0.039\pm0.006$~mag/year during the three years
prior to the supernova, and that this level of variability is 
moderately unusual for other similar stars in M~51.  While there are 
uncertainties about whether the true progenitor was a blue companion to this 
candidate, the result illustrates that there are no technical challenges to 
obtaining fairly high precision light curves of supernova progenitors using 
ground based observations of nearby ($<10$~Mpc) galaxies with wide field 
cameras on 8m-class telescopes.  While other sources of variability may 
dominate, it is even possible to reach into the range of evolution rates 
required by the quasi-static evolution of the stellar envelope.
For M~81, where we have many more epochs and a slightly longer time baseline,
our formal $3\sigma$ sensitivity to slow changes is presently $3$~millimag/year
for a $M_V \simeq -8$~mag star.  In short, there is no observational barrier to
determining whether the variability properties of stars in their last phases of
evolution (post Carbon ignition) are different from earlier phases.
\end{abstract}


\section{Introduction}
\label{sec:introduction}

The last few years have seen steady progress in the identification of the 
progenitors of core-collapse supernovae (ccSNe, see the review by 
\citealt{Smartt2009}). The progenitors of Type~IIP SNe are red supergiants, 
although there is some evidence that the most massive progenitors are less 
massive than the expected upper mass of red supergiants at the ends of their 
lives (\citealt{Kochanek2008}, \citealt{Smartt2009b}, but see
\citealt{Walmswell2011}).  The progenitors of two Type~IIb supernovae have been
identified, a (probably) mass transfer binary system in SN~1993J
(\citealt{Aldering1994}) and the progenitor of SN~2011dh (\citealt{Maund2011},
\citealt{VanDyk2011}).  In two cases, SN~2005gl (\citealt{GalYam2009}) and
SN~1961V (\citealt{Kochanek2011}, \citealt{Smith2011}), the progenitors appear
to be massive, luminous blue variables (LBV).
The dust-enshrouded progenitors of Type IIn SN~2008S (\citealt{Prieto2008b})
and 2008 NGC~300 transient (\citealt{Prieto2008a}) were identified in the
mid-infrared.
No progenitors of Type~Ibc SNe have been identified, and this 
will likely remain challenging due to their lower rates, large expected 
bolometric corrections, and the ease with which a binary companion can be 
optically brighter (\citealt{Kochanek2009}).  Still, progenitor studies are now
well established.

The next frontier is the variability of progenitors.  Very little is known 
observationally about the variability of these stars shortly before explosion.  
The progenitor of the Type IIpec SN~1987A varied by less than a few 
tenths of a magnitude during its last century (see \citealt{Plotkin2004} and 
references therein), while the progenitor of the Type IIb SN~1993J varied by 
less that $0.2$ mag over a 6 month period 9 years before the explosion 
(\citealt{Cohen1995}).  The progenitor of the Type~IIP SN~2008cn was probably 
variable at a level of $\sim 0.2$~mag and the sparse light curve is potentially 
interpretable as an eclipsing binary (\citealt{Elias2009}).  The progenitor of 
the Type~Ib SN~2006jc showed an outburst two years before explosion 
(\citealt{Pastorello2007}). SN~1961V also had an outburst prior to explosion if 
it is interpreted as an SN (\citealt{Kochanek2011}, \citealt{Smith2011}).
In several transients, variability was either observed (SN~2010da,
\citealt{Laskar2010}) or well-constrained (SN~2008S, \citealt{Prieto2008b};
NGC~300-OT, \citealt{Prieto2008a}, \citealt{Thompson2009}) in a dusty stellar
wind rather than directly in the star. 
More generally, outbursts in the last $\sim$century before explosion may
be required to explain the post-explosion evolution of many Type~IIn
SNe (e.g., \citealt{Fox2011}). In short, data are available for very few SNe,
and when it exists it is generally too sparse to interpret.
      
Theoretically, all SN progenitors are variable because their envelopes are 
evolving on a thermal time scale in response to the rapid changes in the core 
luminosity.  In the standard lore, this quasi-equilibrium evolution is too 
small to be observable.  Rough estimates can be extracted from some tabulated 
evolution models, and typical rates in the optical are 
$0.1$-$1.0$~millimag/year (\citealt{Schaller1992}, \citealt{Heger2000a}, 
\citealt{Heger2000b}), but none of these studies was really intended for 
studies of the surface evolution during the last century.  While these rates 
are small, it is certainly possible to achieve the necessary photon counting 
statistics to detect 1~millimag/year changes in progenitors for nearby
($<10$~Mpc) galaxies (see the discussion in \S\ref{sec:discussion}).

It is likely in most cases that other sources of variability will dominate and 
mask the slow evolution of the envelope.  Many massive stars vary either
regularly or irregularly, although good statistics for the evolved stars likely
to be SN progenitors seems to be lacking.  The nature of the variability is
closely related to the stellar type (e.g., \citealt{Szczygiel2010}), but not
as yet in a manner that is a better 
diagnostic of the star than its location in a color magnitude diagram except in 
the limit of helioseismology.  There are theoretical arguments that the 
pulsational properties of red supergiants change with the onset of carbon 
burning (\citealt{Heger1997}), but this has not been developed to the point of 
providing any observational guidance.  Similarly, \cite{Arnett2011} show that 
in three dimensions the nuclear burning fronts of stars in these late phases 
can be very dynamic, which could drive surface effects. Thus, it is likely that 
the progenitors are variable, particularly the red supergiants, but it is 
unknown whether the variability properties of SN progenitors show any 
recognizable difference from stars that have not commenced carbon burning.

Another source of variability comes from shells of material ejected during 
outbursts either as observed for SN~2006jc (\citealt{Pastorello2007}) or 
inferred from the post-SN evolution (e.g., \citealt{Fox2011}).  Ejected material 
has observable effects on the light curve of the progenitor if dust forms in 
the ejected material.  As a dusty shell expands, its optical depth drops as 
$\tau \propto 1/r^2 \propto 1/t^2$ and the star becomes steadily brighter and 
bluer, as is observed for sources such as $\eta$~Carinae (see the review by 
\citealt{Humphreys1994}). Observing these changes, as well as the associated 
mid-IR emission, constrains the time of the eruption if unobserved and the 
amount of material ejected.

Finally, close binary stars can show ellipsoidal variations from a relatively 
broad range of viewing angles and eclipses from a narrow range.  For example, 
the progenitor of SN~1993J had a roughly 15\% probability of producing visible 
eclipses based on the binary evolution models of \cite{Stancliffe2009}.  It 
would, however, require extraordinarily good luck to find a supernova 
progenitor in an eclipsing binary since a more typical probability is 5\% for
a 100\% binary fraction, and for the distance limits we are considering
($<10$~Mpc) the SN rate is only $\sim 1$/year.  In most cases it will likely be
easier to identify candidate binary companions once the SN has faded
(see \citealt{Kochanek2009}).

With wide field cameras on 8.5m class telescopes it is now possible to begin 
exploring these problems in nearby ($<10$~Mpc) galaxies.  While crowding means 
that the Hubble Space Telescope (HST) is generally needed to provide 
photometry, difference imaging methods make it relatively easy to monitor 
individual stars from the ground because luminous variable stars are relatively 
rare and hence not crowded.  This is illustrated in \cite{Gerke2011}, where we 
identified over 100 Cepheids in M~81 using the Large Binocular Cameras (LBC, 
\citealt{Giallongo2008}) on the twin 8.5m Large Binocular Telescope (LBT).  
The shortest period Cepheids in \cite{Gerke2011}, at $P=10$~days, have masses 
of order $M\simeq 6M_\odot$ (e.g., \citealt{Bono2000}) that are well below the 
mass limit for SN progenitors.  We are presently monitoring 25 nearby galaxies 
in the UBVR bands using the LBT/LBC, sparsely monitoring most and more 
intensively monitoring a few.  In addition to the search for variable stars, 
the data also enables two more speculative projects.  The first is to set 
limits on the existence and rate of failed supernovae, massive stars forming 
black holes without a dramatic external signature (\citealt{Kochanek2008}).  
Based on the best current statistics for star formation and supernova rates, 
these could represent up to half of all stellar deaths 
(\citealt{Horiuchi2011}).

The second speculative goal is to study the variability of SN progenitors. 
Unfortunately, the first SN in our sample, SN~2009hd in NGC~3627 
(\citealt{Monard2009}), occurred when we had almost no data and it also lay 
behind a dust lane in the wings of an unobscured bright star (see 
\citealt{Elias2011}).  We obtained no interesting limits on the variability of 
the progenitor, although we should be able to obtain UBVR photometry of the 
progenitor once the SN has faded.  The second SN in our sample is SN~2011dh in 
M~51 (\citealt{Griga2011}).  A candidate progenitor for SN~2011dh was rapidly 
identified by several groups (\citealt{Maund2011}, \citealt{VanDyk2011}).  The 
object is relatively yellow ($T \sim 6000$~K) and there are suggestions that 
the SED may represent a composite of two, presumably binary, stars.  
Furthermore, \cite{Arcavi2011} and \cite{Soderberg2011} argue that the star
which exploded must be a more compact, blue star based on the rapid evolution
of the early-time light curve and spectroscopy.  This would be consistent with
the presence of H$\alpha$ emission in pre-explosion HST images
(\citealt{Szczygiel2011}).
As discussed in \cite{Kochanek2009}, we expect 50-80\% of SNe to occur 
in stellar binaries, and it is relatively easy for the cooler star to dominate 
the optical emission. Combined with the Type~IIb spectroscopic type, the system 
seems very similar to the binary progenitor of SN~1993J. While we have 
yet to carry out the intensive monitoring phase for periodic variables in M~51, 
we have 5/4 epochs of UBV/R data spread over three years and came close to 
observing the star just before it exploded since the next run started on 2 June 
2011 with M~51 as a high priority target.  The candidate progenitor is well 
detected in all epochs and variable.  We will simply refer to this star as the 
progenitor to avoid constant use of the clumsy phrase ``candidate progenitor,'' 
but the ultimate interpretation of its variability clearly depends on resolving 
this ambiguity.  In \S\ref{sec:results} we describe the data and our variability
analysis based on difference imaging.  In \S\ref{sec:discussion} we discuss
some of the implications.
We adopt a distance to M~51 of 8.3 Mpc (\citealt{Poznanski2009}) and
a foreground Galactic extinction of 0.035 mag (\citealt{Schlegel1998}).

\section{Data and Results}
\label{sec:results}

\begin{figure*}
\centerline{\includegraphics[width=6.3in]{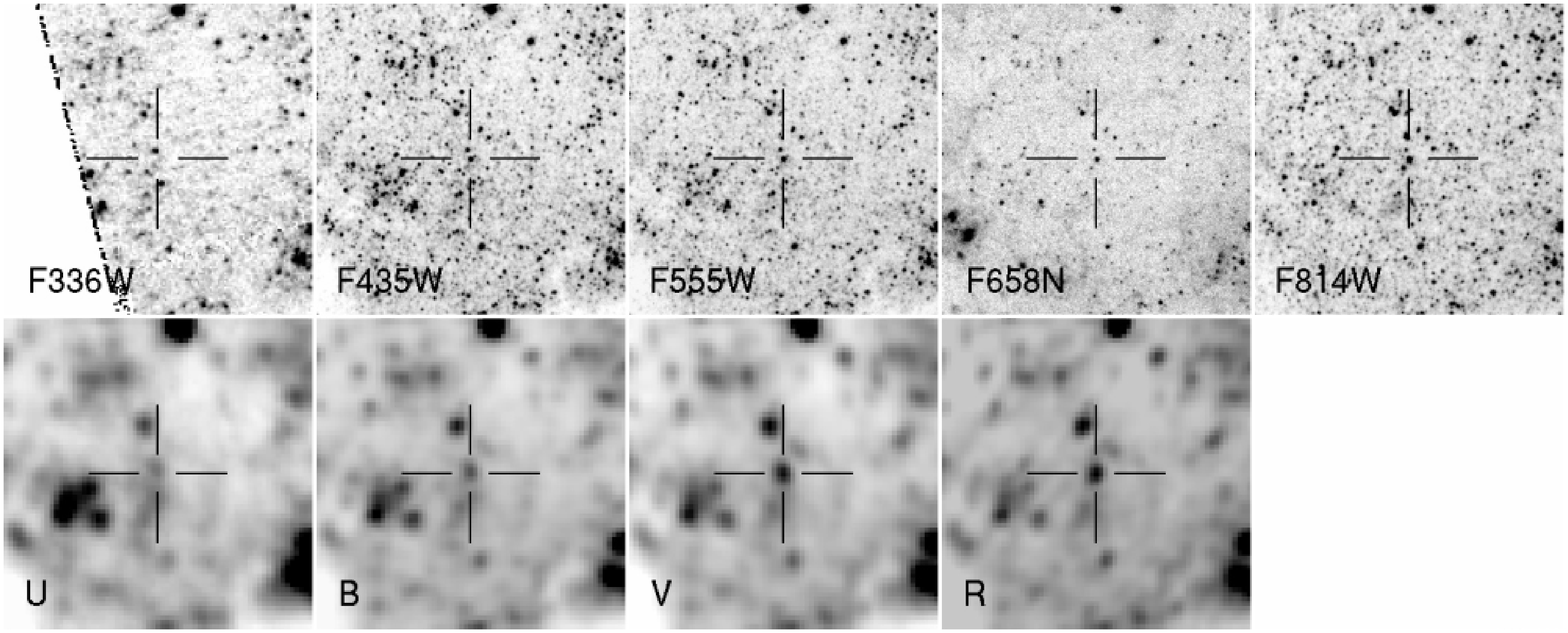}}
\caption{
The HST (top panels) and LBT (bottom panels) pre-explosion images of the region
surrounding the SN 2011dh progenitor
in M~51. The location of the progenitor is marked with a cross hair, the bars
of the cross hair are $2\farcs5$ long and the panels are 15x15 arcsec. Note
the shift in the position of the central object between the R and U-band LBT
images because the U band flux is dominated by the bright blue star (easily
seen in the HST F336W image) to the North-East.
  }
\label{fig:images}
\end{figure*}

We observed M~51 with the LBT/LBC before the SN explosion on 9 March 2008, 28 
January 2009, 19 March 2010, 11 February 2011 and 29 April 2011, and after the 
explosion on 5 and 9 June 2011.  The LBC-Red camera was not available for the 
11 February 2011 epoch.  Table~\ref{tab:obslog} summarizes the observations.  
The LBC cameras have a pixel scale of 0\farcs224, and the galaxy was placed on 
the central chip \#2 of each camera.  The images were bias-corrected and flat 
fielded using sky flats following standard procedures using IRAF mscred tasks.
Fig.~\ref{fig:images} shows the UBVR reference images as compared to the
archival HST images of the region.  The star is clearly visible, but blended
with the blue star 0\farcs5 to the North-East.  The progenitor dominates the
BVR fluxes (based on the HST images, it represents 65\%, 76\% and 87\% of the
B, V and I fluxes, respectively), while the U band flux is dominated by the
blue star and we see a corresponding shift in the location of the peak. Since
the SN progenitor is not visible in the U-band images, we will not discuss them
in great detail.

We analyzed the images using the ISIS difference imaging package 
(\citealt{Alard1998}, \citealt{Alard2000}) with a modified star matching 
procedure based on Sextractor (\citealt{Bertin1996}) that performs more 
reliably in our fields.  The earlier LBC data had some significant rotations 
and translations relative to the nominal field centers, and we discovered that 
the standard ISIS interpolation routine mis-handles interpolation in this 
circumstance. Once identified, the problem was corrected by 
modifying the ISIS spline2.c routine to correctly identify the target column 
associated with each row.  We interpolated the individual sub-images to an 
R-band reference frame using a second order polynomial for the coordinate 
transformation constructed from 100-300 stars with typical rms residuals in the 
stellar matches of less than 0.1~pixels.  A reference image was constructed for 
each band from 6-10 of the sub-images obtained prior to the SN.  We carried out 
the difference imaging both on the individual sub-frames and by combining the 
interpolated sub-frames and then difference imaging the combined images.

\begin{figure*}
\centerline{\includegraphics[width=6in]{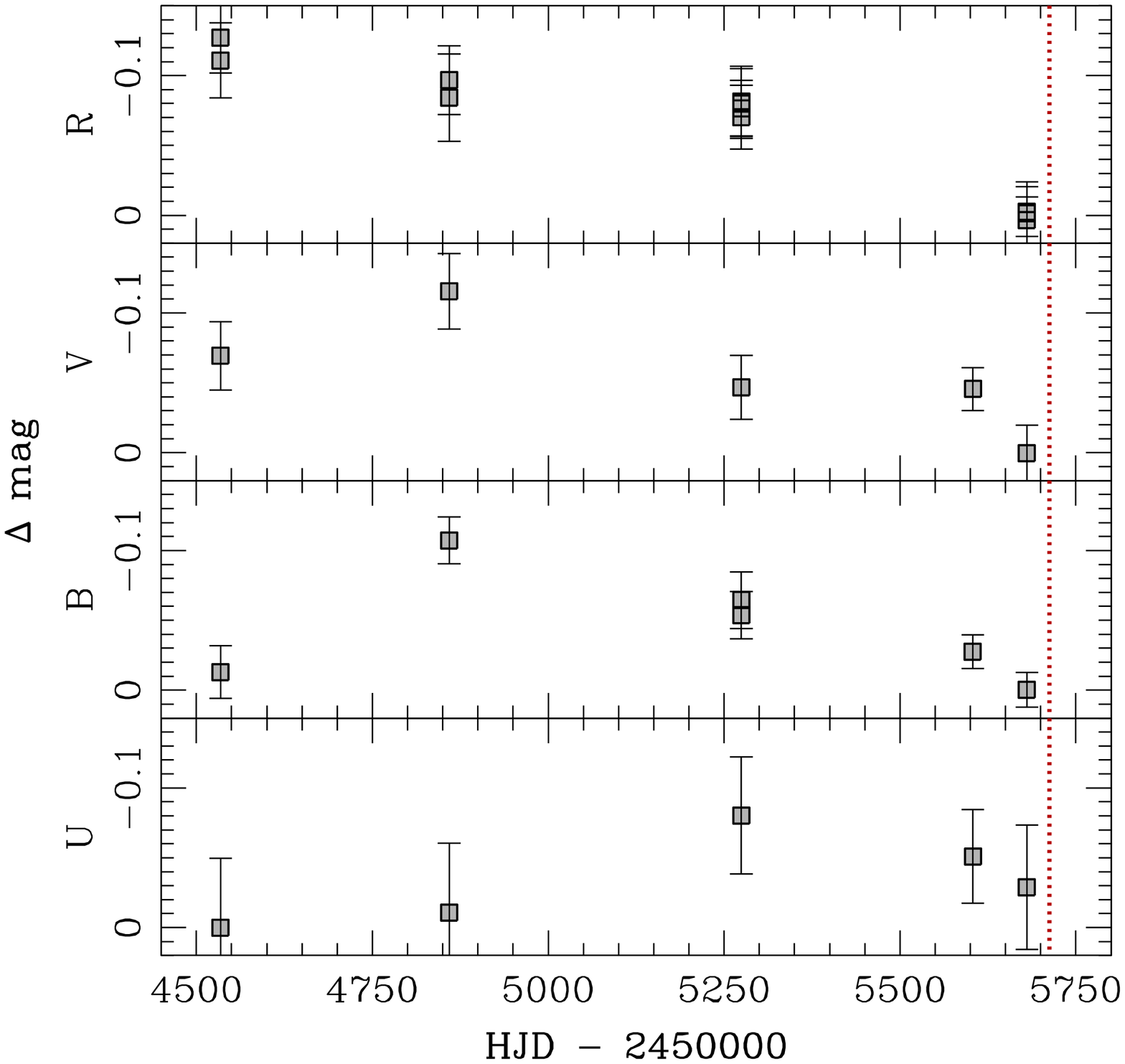}}
\caption{
The U, B, V and R differential light curves of the SN~2011dh progenitor.
The mean magnitude at the epoch of the highest brightness was subtracted from
the light curves.  There is increasing contamination from the nearby blue star
at shorter wavelengths.  The dotted line marks the date of the SN explosion.
}
\label{fig:lc}
\end{figure*}

Fig.~\ref{fig:lc} shows the light curves with the
magnitude at the epoch of highest brightness normalized to zero.  In R band,
the flux is dominated by the progenitor, and over the last three years it faded
by roughly 0.13~mag at a rate of $0.039\pm0.006$~mag/year.  Note that for this
light curve we constructed several stacked images for each epoch rather than
combining all the data into a single image. The results for the separate images
at each epoch are mutually consistent.
The B and V light curves show an initial
rise and then a fall, although an increasing fraction of the light comes from
the contaminating blue star at shorter wavelengths (24\% and V and 35\% at B).
The U band light curve shows no convincing variability and is primarily
emission from the contaminating star.

\begin{figure*}
\centerline{\includegraphics[width=6in]{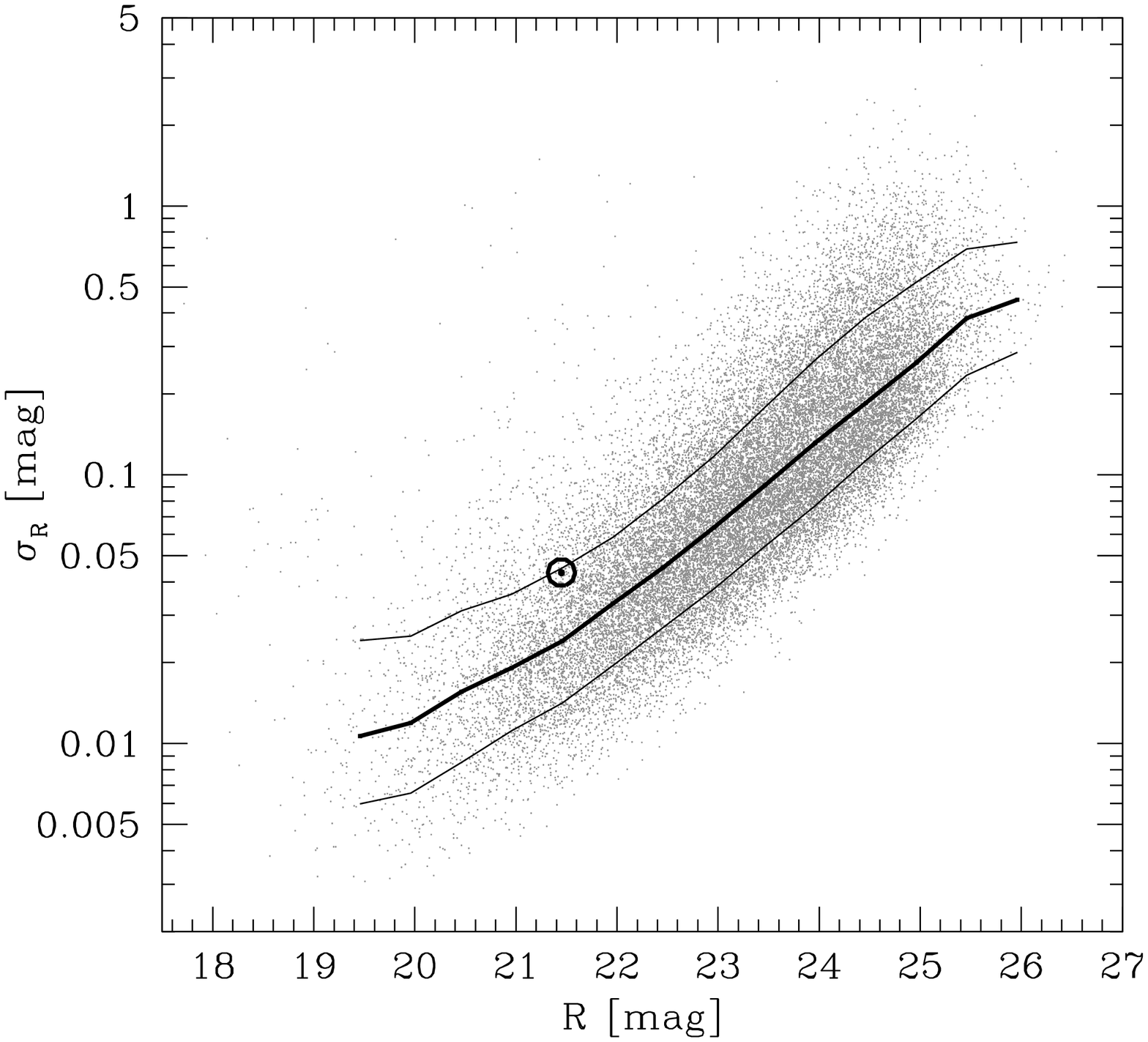}}
\caption{
Variable sources in M~51.  The points show the variance $\sigma_R$ in the
R-band light curves of $\sim 27,000$ sources as a function of the R-band
magnitude.  The heavy line is the median of the $\sigma_R$ distribution within
0.5 mag wide magnitude bins, while the two thin lines symmetrically enclose
68\% of the objects around the median, representing an equivalent of one
standard deviation in this non-gaussian distribution of $\sigma_R$.  The
properties of the SN~2011dh progenitor are marked with an open circle and we
see that it lies just inside the upper thin line, almost falling into the
group of 16\% most variable objects.  We have not carefully inspected this
sample for false sources of variability, so the significance of the progenitor
variability is underestimated.  The general trend simply represents the
scaling of photon counting uncertainties with flux ($\sigma_R \propto R/5$).
}
\label{fig:sigma}
\end{figure*}

We used DAOPHOT (\citealt{Stetson1987}) to catalog the stars on our U, B, V
and R-band ISIS reference images.
We calibrated the U-band data using 6 standard stars from \cite{Pastorello2009} 
and obtaining a calibration uncertainty 0.041~mag. The B, V, and R-band data 
were calibrated using a limited number (14) of Sloan Digital Sky Survey (SDSS 
DR7\footnote{http://cas.sdss.org/dr7/en/tools/search/sql.asp}) stars found on 
chip \#2 (unfortunately, the standards used by \cite{Pastorello2009} are all 
saturated in the LBT BVR images).  We transformed their magnitudes from the 
ugriz to UBVR$_c$ system using \cite{Jordi2006}.  After eliminating stars with 
uncertainties larger than 0.3~mag, we were left with only 7, 5 and 8 stars in 
B, V and R, respectively, leading to calibration uncertainties of 0.230, 0.095 
and 0.081~mag. These absolute calibrations are relatively unimportant for our
present discussion.
We then constructed light curves from the differenced images for 
all the sources in the LBT DAOPHOT catalogs.  Fig.~\ref{fig:sigma} shows the 
variance $\sigma_R$ of these $\sim 27,000$ R-band light curves as a function of 
the R-band magnitude.  The overall trend with magnitude simply represents 
photon-counting statistics and we have made no major effort to clean the 
variability catalog of artifacts.  Since ISIS tends to underestimate the 
uncertainties in light curves, we used Fig.~\ref{fig:sigma} to rescale its 
error estimates.  Given that most sources are not variable, the light curve 
uncertainties should match the median variance seen in Fig.~\ref{fig:sigma}
(heavy line), and this requires scaling the ISIS light curve uncertainties
upwards by a factor of three.  We use these rescaled uncertainties in
Fig.~\ref{fig:lc}. The final light curve of the SN progenitor is presented in
Table~\ref{tab:lightcurve}.  The individual uncertainties are correct for the
light curve, but there are additional global uncertainties from setting the
magnitude of the source in the reference image and the zeropoints that are
reported in the caption. 

\begin{figure*}
\centerline{\includegraphics[width=6in]{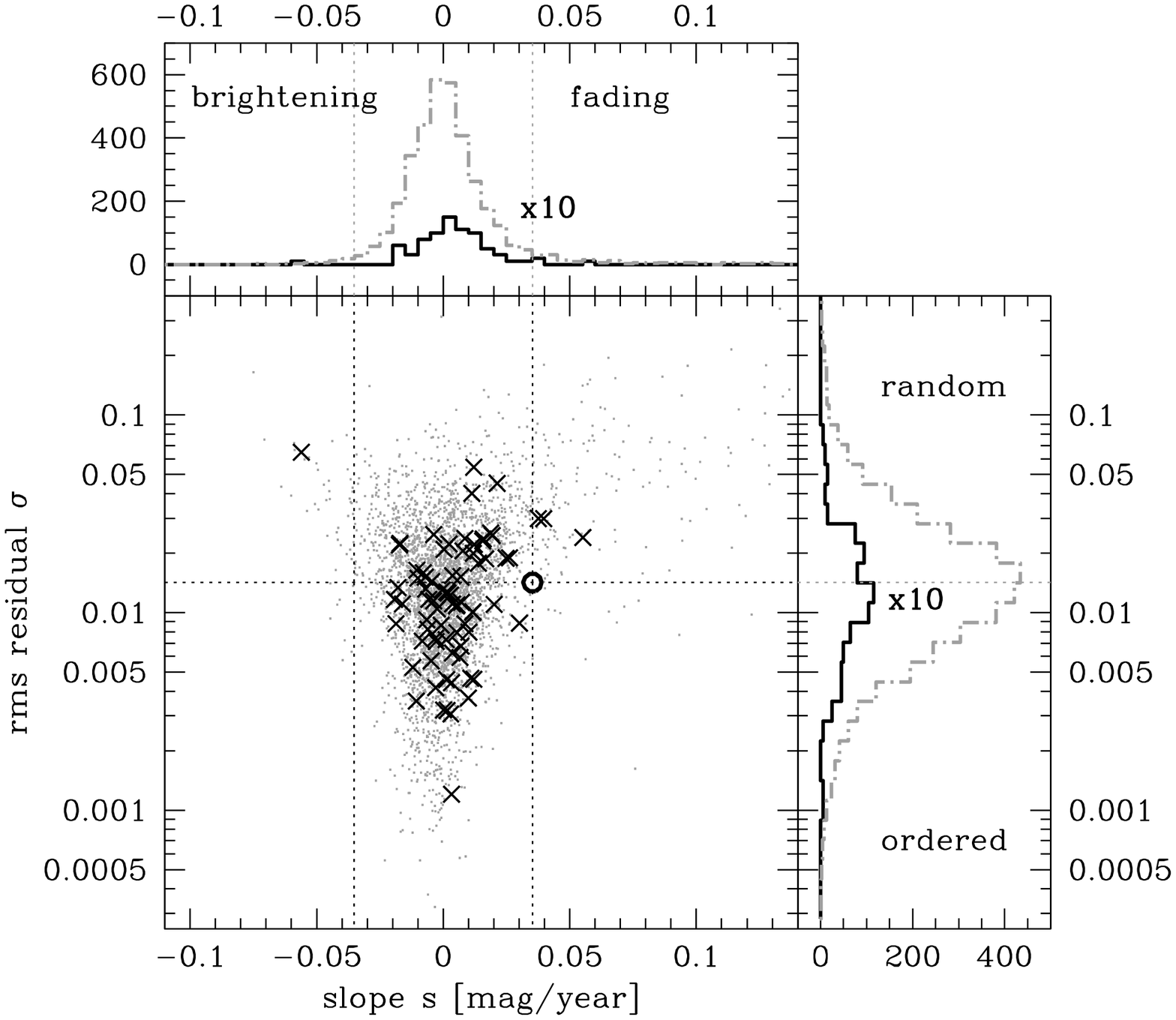}}
\caption{
Variability statistics of M~51 stars.  The scatter diagram shows the 
distribution of objects in light curve slope $s$ and rms residual $\sigma$ for
the progenitor (large circle), for 3800 stars within 1~R~mag of the progenitor
(small points) and for 77 analogue stars of similar luminosity and spectral 
energy distribution (crosses).  The dashed lines mark the values observed for 
the progenitor for both signs of the slope.  The panels to the side show 
projected histograms of the comparison stars, where the histogram for the small 
sample of analogues has been multiplied by ten.  While the rms variability of 
the progenitor is typical of either sample, stars fading as rapidly as the 
progenitor are relatively rare.
}
\label{fig:analogs} 
\end{figure*}

Clearly we cannot interpret such a sparse light curve physically, but we can 
for the first time examine the variability of an SN progenitor as compared to 
other stars.  Keep in mind, however, the debate on the identification of the 
progenitor -- this may be the variability of a binary companion to the actual
SN progenitor.  As shown in Fig.~\ref{fig:sigma}, the progenitor is
significantly more variable than the typical star of its magnitude.  Since its
light curve is roughly a linear decline, we made linear fits $m_i = s t + m_0$
to all the light curves to give a slope $s$ (mag/year, so positive slope is
fading) and the residual dispersion $\sigma$ about the linear fit.
In Fig.~\ref{fig:analogs}.  we show the distribution of the 3800 stars within
$\pm 1$~R~mag of the progenitor in this space of ordered ($s$) and disordered
($\sigma$) variability.  In this variability space, the progenitor properties
are clearly quite different from the typical star, with 93\% of the stars
having smaller slopes in absolute value.  However, note that simply comparing
to stars of similar R band magnitude averages over stars of many types, and
there are probably many false outliers in the distribution because we have not
inspected all the light curves of these objects for artifacts.

We can construct a better comparison sample by searching for ``true'' analogues 
to the progenitor in the B, V, I catalogs obtained from the HST images
(constructed with DAOPHOT).  We matched the magnitudes $m_{a,i}$ of each
potential analogue to those of the progenitor $m_{p,i}$,
\begin{equation}
   \chi^2 = \sum_{i=B,V,I} \left( m_{p,i} - m_{a,i} - \Delta M - R_i \Delta E 
\right)^2 \sigma_0^{-2},
\end{equation}
allowing for a difference in luminosity $|\Delta M|<0.5$~mag and extinction 
$|\Delta E|<0.1$ and accepting those with $\chi^2<4$ for a fixed 
$\sigma_0=0.1$~mag.  This identified 235 such stars, of which 77 lay outside 
the masked regions of the R band LBT image.  Many of the potential analogues 
lie in the central regions of M~51, which are saturated in our R band images, 
while the remainder trace the spiral arms. We visually examined the analogues 
to detect any potential artifacts.  Fig.\ref{fig:analogs} also shows the 
variability properties of these analogue stars (black crosses), and the slope
of the progenitor is still unusually large, with only 5\% (4 objects) of the
analogs showing larger absolute slopes.

\begin{figure*}
\centerline{\includegraphics[width=6in]{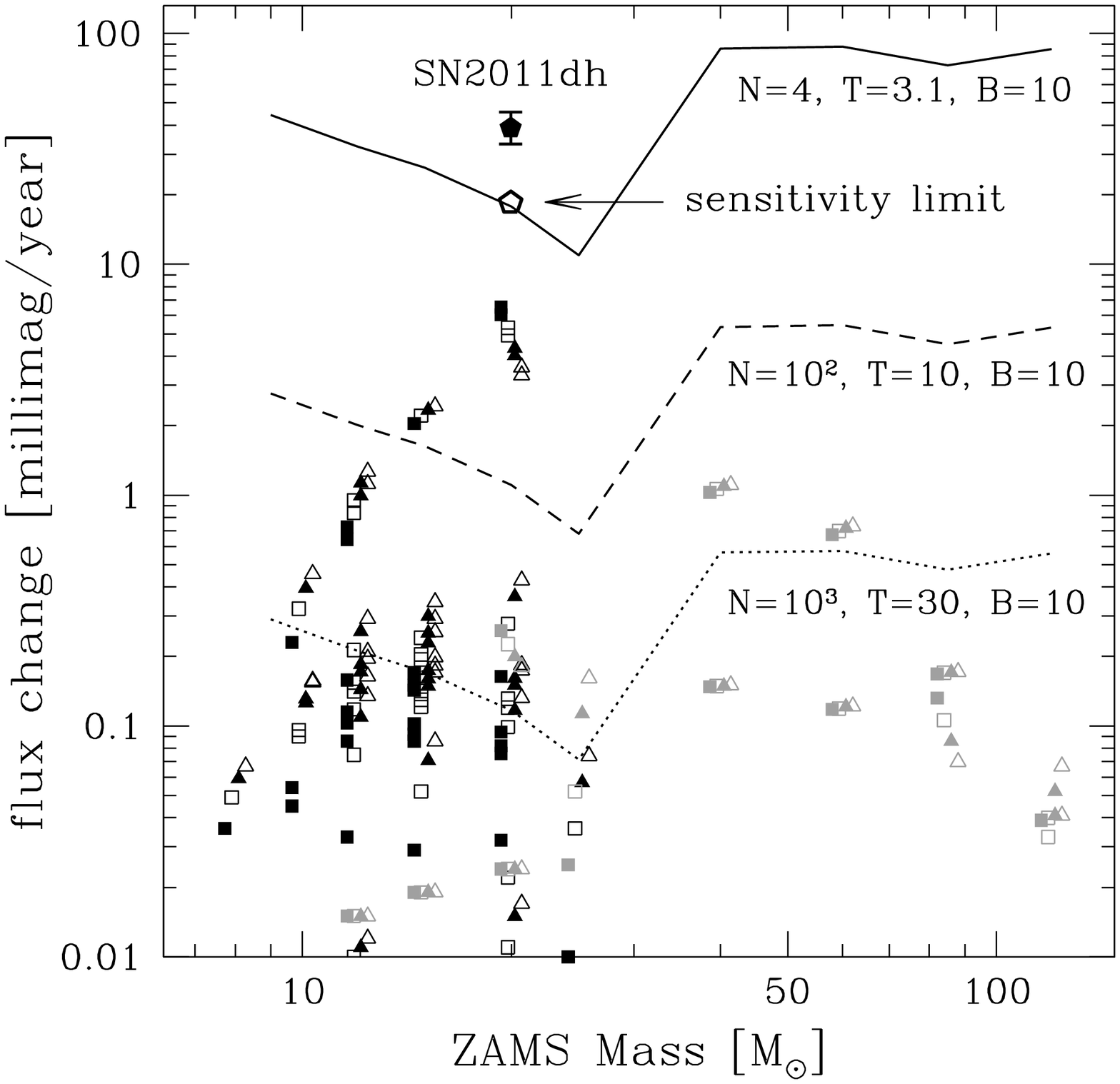}}
\caption{
  Crude estimates of the rate of change in stellar fluxes over the last century 
  before core collapse based on models from \cite{Schaller1992} (gray), 
  \cite{Heger2000a} and \cite{Heger2000b} (black). {it None of these models were,
  however, intended for this purpose.}  A black body is assumed to 
  convert $L$ and $T_e$ into estimates of the rate of change in the B (filled 
  square), V (open square), R (filled triangle) and I (open triangle) 
  magnitudes. The different filters are slightly offset to improve visibility.  
  The filled pentagon shows the slope measured for SN~2011dh and the open 
  pentagon below it shows our estimated sensitivity limit -- the points are 
  placed where the luminosity of the progenitor roughly matches the 
  \cite{Schaller1992} models. The solid curve shows the (statistical) 
  sensitivity of the present data at R-band, assuming $N=4$ epochs spread over 
  $T=3.1$ years, with the background flux in the photometric aperture being 
  $B=10$ times that of the star for an exposure time $t_{exp}=1800$~sec on an 
  8.5~m telescope. The absolute V band magnitudes are estimated from 
  \cite{Schaller1992} and the distance is set to $8.3$~Mpc
  (\citealt{Poznanski2009}).  The dashed curve 
  shows the sensitivity that can be achieved relatively easily from the ground, 
  increasing the baseline to $T=10$~years with $N=10^2$ epochs, and the dotted 
  curve shows a not impossible study with $T=30$~years and $N=10^3$ epochs.
  }
\label{fig:evolve}
\end{figure*}

\section{Discussion}
\label{sec:discussion}

Our primary result is that with only 4/5 epochs of ground based data there is
no difficulty in detecting low level variability in a supernova progenitor 
candidate at a distance of almost 10~Mpc.  In the R-band we find a relatively 
steady decline of $s=0.039 \pm 0.006$~mag/year over a three year baseline with 
rms residuals of only $0.02$~mag.  With so few epochs, we cannot interpret this 
physically and the uncertainty about the true identity of the progenitor 
complicates the interpretation in any case.  As noted by \cite{Kochanek2009}, 
50-80\% of ccSNe should be in stellar binaries at the time of explosion.  
Furthermore, if the binary consists of a blue, hot star and a red (in this case 
yellow) cool star and the blue star explodes, it will not be uncommon for the 
visual emission from the progenitor to be dominated by the red companion rather 
than the star which exploded. This issue will be resolved as the direct 
emission from the SN fades, although we should note that the system most 
advanced as a possible analogue, SN~1993J, produces so much emission from the 
expanding shock moving through the circumstellar medium that the binary 
companion only became observable a decade later (see \citealt{Maund2004}).
  
While the utility of finding that the progenitor is an eclipsing binary is 
obvious, one could legitimately ask whether detecting other sources of 
variability has any use.  At a very basic level, these stars are different from 
all other stars in their galaxies.  In a Hertzsprung-Russel diagram this is not 
apparent because there is nothing special about their luminosities and surface 
temperatures.  Variability opens a new window to search for these differences.  
There may be none, but the default expectation that there are no precursor 
signals to SNe is essentially based on assuming that the only effect is the 
quasi-static evolution of the stars and that this is too slow to detect.

How slow is the quasi-static evolution?  Fig.~\ref{fig:evolve} shows estimates 
of the rate of evolution for the last century before collapse derived from the 
models tabulated by \cite{Schaller1992}, \cite{Heger2000a} and 
\cite{Heger2000b}.  The stars are generally becoming slightly fainter and 
hotter at rates of 0.1-1~millimag/year.  Bear in mind that these models were 
not intended for this purpose.  Based on photon statistics, it is possible to 
detect such slow rates of change.  Given $N$ observations with an 8.5~m 
telescope, uniformly spaced over time $T$, with exposure time $t_{exp}$ for 
each epoch, the $3\sigma$ sensitivity to a temporal gradient is 
\begin{equation}
   s_3 \simeq { 0.040~\hbox{mag} \over T }
     \left( { D \over 10~\hbox{Mpc} } \right)
     \left( { (1+B) \hbox{ksec} \over N t_{exp} }\right)^{1/2}
      10^{-0.2(M_V+8)}
     \label{eqn:signal}
\end{equation}
for a star of absolute luminosity $M_V$ at distance $D$ with the background 
flux in the extraction aperture being $B$ times that of the star.  For the 
R-band observations of M~51, we have $T=3.1$~years, $N=4$, $Nt_{exp}=8000$~sec, 
$D=8.3$~Mpc, and $B \simeq 10$, yielding $s_3 \simeq 14$~millimag/year for the 
$M_V \simeq -8$~mag progenitor star.  This result does not depend significantly
on the adopted SN distance -- if we take $D=7.1$~Mpc (used by \citealt{Maund2011})
or $D=7.66$~Mpc (\citealt{VanDyk2011}), then $s_3 \simeq 12-13$~millimag/year.
The formal estimate from our progenitor 
light curve is that we reached $s_3 \simeq 18$~millimag/year, roughly 
consistent with this estimate but well above even the highest rates shown in 
Fig.~\ref{fig:evolve}. 

Consider, however, the data we have accumulated on M~81 to study its variable 
stars (\citealt{Gerke2011}).  With $D=3.6$~Mpc, $N=50$, $N t_{exp}=21540$~sec, 
and $T=4.1$~years, the nominal $3\sigma$ sensitivity of $s_3=3$~millimag/year 
(at $M_V=-8$) is below some of the quasi-static evolution rate predictions.
Extending the time baseline to $10$~years reaches $s_3 \simeq 1$~millimag/year,
and campaigns with $N=1000$ and $T=30$~years yielding
 $s_3 \simeq 0.1$~millimag/year are conceivable.  Thus, it is possible to
obtain the photon statistics needed to probe this phase of stellar evolution,
and we know from ground-based milli-magnitude photometry of planetary transits
(see the review by \citealt{Winn2010}) and $\sim 10^{-2}$~milli-magnitude
photometry of transits in space (\citealt{Borucki2009}) that systematic errors
can be controlled well enough to approach the statistical limits.
A milli-magnitude is not what it used to be.
A few other regimes of stellar evolution, such as the post-helium flash
evolution of stars onto the horizontal branch (Bildsten, private comm.) and
some massive stars in the Hertzsprung gap should also be fast enough to
observe, albeit not at Mpc distances. We do need theoretical models intended
for making estimates of the surface evolution in these last phases,
particularly since most studies of stellar evolution literally freeze the
envelope for the last phases.
  
The systematic problem that will most limit measurements of quasi-static 
evolution is that stars vary on many time scales other than this evolutionary 
time scale.  Variability acts like an added source of noise, essentially adding 
$ B \simeq (A/\sigma_{phot})^2$ to Eqn.~\ref{eqn:signal}, where $A$ is the 
amplitude of any unmodeled variability and $\sigma_{phot}$ is the typical 
photometric error at any epoch.  If other sources of variability dominate, then 
the question becomes whether the variability of progenitors can be 
distinguished from that of other stars.  With the sparse data we have, we can 
only consider simple metrics.  Here we examined variability in the space of a 
mean trend and the residuals around it, finding that the progenitor has 
modestly unusual variability properties.  Since there is essentially no 
theoretical guidance on variability in these late phases, other than the study 
by \cite{Heger1997} that the oscillation properties change, and we are faced 
with the additional uncertainties about the nature of the progenitor, it is 
premature to draw conclusions.  We will probably need variability statistics on 
several progenitors to begin having a clear path for interpreting the results.

The particular properties of one ambiguous object are not a revolution.  But 
the ability to make the measurements may be revolutionary. {\it Using 
difference imaging techniques we can measure the variability of any SN 
progenitor within $10$~Mpc from a ground-based 8.5m telescope at levels that 
certainly approach and may reach the variability expected from the quasi-static 
evolution of the stars. } Any variability significantly above that level is 
trivial, and we simply face the quantitative question of whether the 
variability of post-Carbon burning stars can be distinguished from that of 
other stars.
In other words, we could ask whether the variability properties can be
used to point at stars about to explode.
The ``about to'' is likely on the
order of the present duration of human civilization (the $10^3-10^4$~years 
after Carbon ignition), but this is still a remarkably narrow window compared 
to the life times of even massive stars.  The important point about our 
observations of the progenitor of SN~2011dh is that these are now observational 
questions -- we do not know if the answer will be boring or exciting, but we 
know we can answer the question.
 
Forty galaxies produce 90\% of the local ($<10$~Mpc) ccSNe rate of roughly 
1~SN/year (see \citealt{Kochanek2008}).  It conservatively requires 4 nights on 
an 8m~class telescope with a wide field camera (LBC on LBT, Suprimecam on 
Subaru, or to a lesser degree, IMACS/Megacam on Magellan) to provide one epoch 
of data with depth comparable to our present data for all 40 galaxies.  Such 
single epoch data generally exists but is not very useful because seeing 
induced confusion means that photometry of individual stars is essentially 
impossible at these distances with ground-based optical resolution.  Accurate 
single epoch fluxes require the high resolution of the Hubble Space Telescope 
(HST).  Galaxies at these distances are not, however, crowded with luminous 
variable stars even at ground-based resolutions, and, as shown in our study of 
Cepheids in M~81 (\citealt{Gerke2011}), variability should be measured from the 
ground with difference imaging and only the absolute calibrations done from 
space.  Getting to the modest numbers of epochs we use here is relatively easy, 
roughly 20 nights to obtain 5 epochs for every galaxy.  This is approximately 
where we stand in our LBT survey of 25 of these 40 galaxies - we have a median 
of 5 epochs.  Reaching 30 epochs is expensive, but this is roughly the 
threshold where one can identify and phase periodic variables like Cepheids and 
build long term light curves of fainter transients.  At least in our LBT 
survey, we are trying to reach this level for a subset of the galaxies that are 
interesting for studies of the distance scale (e.g., M~81, NGC~4258, M~101) or 
where there are interesting, faint transients to be monitored (e.g., SN~2008S
in NGC~6946, see \cite{Kochanek2011}).  Achieving the next level, 100 epochs, 
probably requires a dedicated imaging telescope like LSST, but only represents 
5-10\% of the observing time over a period of 20 years.  With this many epochs, 
most eclipsing binaries will be identified and it should also enable searches 
for microlensing events.

\acknowledgements

We thank T.~A. Thompson for comments on the manuscript, and the observers who
took LBT data used in this paper:
J. Antognini,
D. Atlee,
R. Beaton,
A. Bedregal,
J. Blackburne,
M. Dietrich,
P. Garnavich,
J. Johnson,
P. Martini,
R. Pogge,
J. Prieto,
G. Privon,
K. Rueff,
and
L. Watson.
The authors are supported by NSF grant AST-0908816.
Based in part on observations made with the Large Binocular Telescope. The LBT 
is an international collaboration among institutions in the United States, 
Italy and Germany. The LBT Corporation partners are: the University of Arizona 
on behalf of the Arizona university system; the Istituto Nazionale di 
Astrofisica, Italy; the LBT Beteiligungsgesellschaft, Germany, representing the 
Max Planck Society, the Astrophysical Institute Potsdam, and Heidelberg 
University; the Ohio State University; and the Research Corporation, on behalf 
of the University of Notre Dame, University of Minnesota and University of 
Virginia.
This work is based in part on observations made with the NASA/ESA 
Hubble Space Telescope, obtained from the data archive at the Space Telescope 
Institute. STScI is operated by the association of Universities for Research in 
Astronomy, Inc. under the NASA contract NAS 5-26555. This research has made use 
of the NASA/IPAC Extragalactic Database (NED) which is operated by the Jet 
Propulsion Laboratory, California Institute of Technology, under contract with 
the National Aeronautics and Space Administration.

{\it Facilities:}  \facility{Large Binocular Telescope, HST}

\begin{deluxetable}{rrc}
\tablecaption{Observation Log}
\tablewidth{0pt}
\tablehead{
\multicolumn{1}{c}{Date} &\multicolumn{1}{c}{LBC Red} &\multicolumn{1}{c}{LBC Blue}  \\
               &\multicolumn{1}{c}{$N_R \times t_{exp}$} &\multicolumn{1}{c}{$N_V,N_B,N_U \times t_{exp}$} }
\startdata
9 March 2008      &$ 6  \times 300$  &$2,2,2 \times 300$ \\
28 January 2009   &$ 8  \times 200$  &$3,2,3 \times 200$ \\
19 March 2010     &$ 14 \times 200$  &$3,6,5 \times 200$ \\
11 February 2011  &                  &$3,3,3 \times 200$ \\
29 April 2011     &$ 9 \times 200$   &$3,3,3 \times 200$ \\
5 June 2011       &$ 9 \times 200$   &$3,3,3 \times 200$ \\
9 June 2011       &$ 9 \times 200$   &$3,3,3 \times 200$ \\
\enddata
\tablecomments{ Exposure times are in seconds. }
\label{tab:obslog}
\end{deluxetable}

\begin{deluxetable}{lccccccccc}
\tablecaption{The progenitor light curve}
\tablewidth{0pt}
\tablehead{
\colhead{Date} &\colhead{HJD-2450000} &\colhead{U} &\colhead{$\sigma_{U}$} &\colhead{B} &\colhead{$\sigma_{B}$} &\colhead{V} &\colhead{$\sigma_{V}$} &\colhead{R} &\colhead{$\sigma_{R}$}    \\
&\colhead{[days]}      &\colhead{[mag]}  &\colhead{[mag]}   &\colhead{[mag]}   &\colhead{[mag]}  &\colhead{[mag]}  &\colhead{[mag]}  &\colhead{[mag]}  &\colhead{[mag]}  }
\startdata
9 March 2008      &54534.5 & 21.796 & 0.050 & 21.908 & 0.019 & 21.275 & 0.024 & 21.322 & 0.025 \\
                  &        & \nodata&\nodata& \nodata&\nodata& \nodata&\nodata& 21.338 & 0.027 \\
28 January 2009   &54859.4 & 21.785 & 0.050 & 21.814 & 0.017 & 21.228 & 0.027 & 21.352 & 0.024 \\
                  &        & \nodata&\nodata& \nodata&\nodata& \nodata&\nodata& 21.365 & 0.031 \\
19 March 2010     &55274.5 & 21.716 & 0.042 & 21.857 & 0.020 & 21.297 & 0.023 & 21.368 & 0.025 \\
                  &        & \nodata&\nodata& 21.867 & 0.017 & \nodata&\nodata& 21.369 & 0.025 \\
                  &        & \nodata&\nodata& \nodata&\nodata& \nodata&\nodata& 21.379 & 0.023 \\
                  &        & \nodata&\nodata& \nodata&\nodata& \nodata&\nodata& 21.372 & 0.020 \\
11 February 2011  &55603.5 & 21.745 & 0.033 & 21.893 & 0.012 & 21.298 & 0.015 & \nodata&\nodata\\
29 April 2011     &55680.4 & 21.767 & 0.044 & 21.921 & 0.012 & 21.344 & 0.020 & 21.446 & 0.018 \\
                  &        & \nodata&\nodata& \nodata&\nodata& \nodata&\nodata& 21.448 & 0.023 \\
                  &        & \nodata&\nodata& \nodata&\nodata& \nodata&\nodata& 21.452 & 0.017 \\ 
\enddata
\tablecomments{
 The uncertainties on the light curves do not include the uncertainties in the
 magnitude of the star on the reference image.  The DAOPHOT uncertainties in 
 the reference image magnitude are $0.033$ in U, $0.051$ in B, $0.036$ in V, 
 and $0.026$ in R, and the magnitude calibration uncertainties are $0.100$ in
 U, $0.230$ in B, $0.095$ in V and $0.081$ in R.
 }
\label{tab:lightcurve}
\end{deluxetable}

\end{document}